\documentclass[]{jfm}

\usepackage{multirow}
\usepackage{subfigure}
\usepackage{graphicx}
\usepackage{newtxtext}
\usepackage{newtxmath}
\usepackage{natbib}
\usepackage{hyperref}

\usepackage{amsmath}

\makeatletter
\newcommand{\specialnumber}[1]{%
	\def\tagform@##1{\maketag@@@{(\ignorespaces##1\unskip\@@italiccorr#1)}}}
\newcommand{\specialeqref}[2]{\begingroup
	\def\tagform@##1{\maketag@@@{(\ignorespaces##1\unskip\@@italiccorr#2)}}%
	\eqref{#1}\endgroup}

\newcommand{\rmd}{\mathrm{d}}

\newcommand{\bk}{\mathbf{k}}
\newcommand{\bx}{\mathbf{x}}
\newcommand{\bc}{\mathbf{c}}
\newcommand{\be}{\mathbf{e}}
\newcommand{\br}{\mathbf{r}}
\newcommand{\bt}{\mathbf{t}}
\newcommand{\bU}{\mathbf{U}}
\newcommand{\bu}{\mathbf{u}}

\newcommand{\mo}{\mathcal{O}}
\newcommand{\parallelsum}{\mathbin{\!/\mkern-5mu/\!}}

\title{Coastal wave refraction in variable currents over a varying bathymetry}

\author{Trygve Halsne 
        \aff{1}\thanks{Email address for correspondence: trygve.halsne@met.no}
        and Yan Li \aff{2} 
        }
\affiliation{
\aff{1}Norwegian Meteorological Institute, N-5007 Bergen, Norway

\aff{2}Department of Mathematics, University of Bergen, N-5007 Bergen, Norway}
\date{}

\begin{document}

\newcommand*{\g}{\text{\slshape g}}
\newcommand*{\kk}{\mathbf{\hat{k}}}
\newcommand*{\ttt}{\mathbf{\hat{t}}}
\newcommand*{\uu}{\mathbf{u}}
\newcommand*{\rd}{\dot{\mathbf{r}}}
\newcommand*{\nn}{\mathbf{\hat{n}}}

\maketitle

\section*{Abstract}
Refraction is the predominant mechanism causing spatially inhomogeneous surface gravity wave fields. 
However, the complex interplay between depth- and current-induced wave refraction remains poorly understood.
Assuming weak currents and slowly varying bathymetry, we derive an analytical approximation to the wave ray curvature, which is validated by an open-source ray tracing framework. 
The approximation has the form of linear superposition of a current- and depth-induced component, each depending on the gradients in the ambient fields.  
This separation enables quantification of their individual and combined contributions to refraction.
Through analysis of a few limiting cases, we demonstrate how the sign and magnitude of these components influence the wave refraction, and identify conditions where they either amplify or counteract each other.
We also identify which of the two plays a dominant role.
These findings provide physically resolved insights into the influence of current- and depth-gradients on wave propagation, and are relevant for applications related to remote sensing and coastal wave forecasting services. 

\section{Introduction}
Refraction is a key mechanism modulating the surface gravity wave field.
For example, the horizontal wave field variability in deep waters is dictated by current-induced refraction \citep{ardhuin_small-scale_2017,boas2020wave}---a mechanism that also influences the occurrence probability of so-called freak waves \citep{hjelmervik_freak_2009,onorato11,smith_giant_1976, white_chance_1998}. 
Most human marine activities take place in the coastal or near-coastal zone.
Such regions are often classified as intermediate or shallow with respect to a characteristic wavelength. 
As a consequence, the wave propagation is typically accompanied by variable currents and varying bathymetry, which can give rise to hazardous sea states \citep{halsne_resolving_2022,li23review,zheng23}. 

Under the geometric optics approximation, wave action density propagates along wave rays \citep{whitham_general_1965}. 
An appropriate measure of wave refraction is the ray curvature, which in plain language means the departure from a straight line.
Approximate ray curvature solutions have been derived for conditions with either depth- or current-induced refraction \citep{ arthur_direct_1952,dysthe_refraction_2001, kenyon_wave_1971}. 
These have provided valuable insights into where wave refraction becomes important and where to expect modulations, both local and non-local, of the wave field \citep[e.g.,][]
{boas2020wave, gallet_refraction_2014, quilfen2018storm}.
Under mixed conditions, when both a varying current and bathymetry are present, the analysis is typically carried out through direct numerical integration of the wave ray equations \citep[e.g.,][]{arthur_refraction_1950, romero_submesoscale_2020, halsne_resolving_2022, jonsson_current-depth_1980}, where it remains physically unresolved as to how the two play a joint role in the wave refraction.
The purpose of this work is to develop a theoretical model that elucidates this interplay.

Knowledge about where, and under which conditions, one might expect strong wave-current-depth interactions is considered valuable information for various geophysical applications. 
For instance, in recent decades there has been an increased focus on retrieving current or bathymetry information using remote sensing observations of the spatially varying wave field \citep[see e.g.,][]{hessner14,klotz2024nearshore,lenain2023airborne,lund15,smeltzer19,stewart_hf_1974}.
Furthermore, recent work shows that a prescribed mean flow field can be used to map the horizontal wave height variability in deep waters, by taking into account the directional diffusion of wave action caused by current-induced refraction \citep{boas_directional_2020, smit2019swell,wang_scattering_2023}.
In water regions where both a varying current and bathymetry are present, the development of remote sensing techniques and simplified prediction models is based on an explicit representation of how the current and bathymetry play a role in modulating the wave field.
Nevertheless, to the best of the authors' knowledge, such an explicit model has not been developed yet.

Based on the previous discussion, it is understood that it remains an open research question as to how wave fields are altered by their ambient environments in the presence of both a varying current and bathymetry. 
Hence, the main objective of this work is to address this question by developing a simplified theoretical model accounting for the combined effects of varying current and bathymetry on wave refraction. 
This paper is laid out as follows: 
We first derive an analytical approximation to the curvature of wave rays in \S~\ref{sec:theory} under the weak current assumption and geometric optics approximation.
Then, the complex interplay between current- and depth-induced refraction is explored in a few limiting cases, by utilizing an open-source ray tracing framework (\S~\ref{sec:evaluation}). Finally, the conclusions are drawn in \S~\ref{sec:conclusions}. 

\section{Wave ray theory}\label{sec:theory}
We consider three-dimensional surface waves atop a background current and a varying bathymetry, assuming incompressible and inviscid flows. A still water surface is considered at the vertical axis $z=0$ in the horizontal $x$--$y$ plane. Let $\bU=[U_1(\bx),U_2(\bx)]$ be the velocity vector of the background current in the horizontal plane with $\bx=(x,y)$ the position vector, $U_1$ and $U_2$ the velocity component in the $x$- and $y$-direction, respectively. The velocity vector $\bU$ is assumed to be depth uniform, and slowly dependent on the time $t$, and the position vector $\bx$ compared with the rapidly varying phase of the surface waves. A water depth $h(\bx)$, is considered, which varies mildly in the horizontal plane. 

Without the loss of generality, we let $\theta(\bx,t) = \bk(\bx,t)\cdot\bx - \omega(\bx,t) t$ be
the spatial-temporal rapidly varying phase of waves 
such that
\begin{equation}
    \label{eq:k_def}
        \nabla \theta = \bk,
        ~ 
        \partial_t\theta = -\omega,
         \text{ and }
        \partial_t\bk(\bx,t) + \nabla \omega(\bx,t) = 0 
        \specialnumber{a,b,c}
\end{equation}
by definition. 
Here, $\nabla=(\partial_x,\partial_y)$; 
$\bk$ and  $\omega$ denote the local wave vector and angular frequency of waves, which are modified by the presence of both current and varying bathymetry, obeying the  dispersion relation as follows \citep{peregrine_interaction_1976}
\begin{align}
   \label{eq:dispR}
    \omega(\bk,\bx) = \bk\cdot\bU +\Omega(\bk,\bx), 
\end{align}
where $\Omega(\bk,\bx) =\sqrt{gk\tanh kh(\bx)}$ denotes the intrinsic wave frequency in the absence of currents, and $k=|\bk|$ denotes the magnitude of the wave vector. For convenience and later reference, we introduce the group velocity vector and phase velocity associated with the intrinsic wave frequency as follows, respectively,
\begin{equation}
    \bc_{g,i}(\bk,\bx) = \nabla_k\Omega
    \text{ and } c(\bk,\bx) = {\Omega}/{k}, 
    \specialnumber{a,b}
\end{equation}
where the operator $\nabla_k = (\partial_{k_x},\partial_{k_y})$ denotes the gradient in the $\bk$ space.

\subsection{Rays and their unit tangent vector}
Introduce $\br(t)= (x_r(t),y_r(t))$ to denote the position vector of the rays in the horizontal plane which are time dependent. According to the definition \citep[their \S 3.6]{mei05},
\begin{equation} \label{eq:raydef}
        \dfrac{\rmd \br}{\rmd t} 
        = \nabla_k\omega(\bk,\bx)
        \equiv \bU + \bc_{g,i}
        \quad\to \quad
        \dfrac{\rmd y_r}{\rmd x_r} 
        = \dfrac{\dot{y}_r}{\dot{x}_r},
        \specialnumber{a,b}
\end{equation}
where the dot denotes the derivative with respect to $t$. Here, \specialeqref{eq:raydef}{a} denotes the absolute group velocity vector and \specialeqref{eq:raydef}{b} leads to the general expression for the local slope of the rays which will be used for the ray curvature presented in \S~\ref{sub:raycurv}. The definition of the rays has the physical meaning of wave group trajectories, i.e., the rays are locally parallel to the absolute group velocity everywhere at all times, and denote the direction of wave action propagation \citep{whitham_general_1965}. 

The unit tangent vector of the rays $\bt$, is essential to the explicit expression of the ray curvature, which can be obtained by noting that $(\rmd \br/\rmd t) \parallelsum \bt$ and $|\bt|=1$, suggesting the following identities hold
\begin{equation} \label{eq:bt_def}
    \bt = \dfrac{\bU+\bc_{g,i}}{|\bU+\bc_{g,i}|}
    \text{ and }
    \bt = \left[ 
    \dfrac{\dot{x}_r}{|\dot{\br}|}, \quad 
    \dfrac{\dot{y}_r}{|\dot{\br}|}
           \right]. 
    \specialnumber{a,b}
\end{equation}
Here, both the expressions will be used for the derivation of the ray curvature. The assumption of weak current compared with the group velocity of the waves is translated to the definition of the dimensionless current velocity as follows 
\begin{equation}
    \bu=[u,v] ={\bU}/{c_{g,i}}
\end{equation}
such that $\mo(|\bu|)\sim \varepsilon$,
with $c_{g,i}=|\bc_{g,i}|$, $u$ and $v$ representing the  component of $\bu$ in the $x$- and $y$-direction, respectively, and $\varepsilon$ denoting a small dimensionless scaling parameter.
Thus, the unit tangent vector given by \specialeqref{eq:bt_def}{a} can be expressed as
\begin{align}\label{eq:bt_exp}
    \bt = \dfrac{\bu+\be_k}{|\bu+\be_k|} = 
     \be_k + \bu -(\bu\cdot\be_k)\be_k + \mo(\varepsilon^2),
\end{align}
where $\be_k=\bk/k$ denotes the unit vector in the same direction as the local wave vector. The approximation to the unit tangent vector given by \eqref{eq:bt_exp} is identical to \citet[expression (3)]{dysthe_refraction_2001} for the limiting cases of deepwater waves. 

\subsection{The ray curvature}
\label{sub:raycurv}
Recall that the curvature $\kappa$, of rays can be expressed in a parametric form as follows \citep[e.g.,][]{mathiesen_wave_1987}
\begin{align}\label{eq:kpdef}
    \kappa = 
    \dfrac{\left|\dot{x}_r\ddot{y}_r - \dot{y}_r\ddot{x}_r \right|}
    {\left[
    (\dot{x}_r)^2 + 
    (\dot{y}_r)^2
    \right]^{3/2}}
    \equiv 
    \left| \bt \cdot \dfrac{[\ddot{y}_r,  - \ddot{x}_r ] }{(\dot{x}_r)^2 + 
    (\dot{y}_r)^2}
    \right|
    ,
\end{align}
where the double dots denotes the second-order derivative with respect to time. It is now clearly seen in \eqref{eq:kpdef} that the curvature relies on an explicit expression for both the unit tangent vector of the rays and the second derivative of the ray trajectories with respect to time. The latter can be obtained by definition:
\begin{equation}
    \label{eq:d2rdt}
    \dfrac{\rmd^2 \br}{\rmd t^2} = [\ddot{x}_r, \ddot{y}_r] 
    \text{ or }
     \dfrac{\rmd^2 \br}{\rmd t^2}  =~  \dfrac{\rmd}{\rmd t}
      (\bU + \bc_{g,i}),
      \specialnumber{a,b}
\end{equation}
the latter of which should be evaluated on the time-dependent rays, i.e., $\bx=\br(t)$ should be noted for \specialeqref{eq:d2rdt}{b}. Thereby, we arrive at 
\begin{align}\label{eq:d2rdt_exp}
    \dfrac{\rmd^2 \br}{\rmd t^2}
    =~  
      (\dfrac{\rmd \bx}{\rmd t}\cdot\nabla )\bU
      + \left(\dfrac{\rmd \bx}{\rmd t}\cdot\nabla\right)\bc_{g,i}(\bk,\bx)
       +
      \left(\dfrac{\rmd \bk}{\rmd t}\cdot\nabla_k\right)\bc_{g,i}(\bk,\bx),
\end{align}
for $\bx=\br(t)$, where $\rmd/\rmd t$ denotes the material derivative with respective to time. We repeat that expression \eqref{eq:d2rdt_exp} is obtained by applying the quasi-stationary background current assumption $\p_t\bU\simeq\mathbf{0}$, meaning that the background current $\bU$ is slowly varying in time compared with the rapidly varying phase and rays. The terms on the right hand side of \eqref{eq:d2rdt_exp} can be otherwise readily evaluated but the material derivative of the wave vector $\rmd \bk/\rmd t$, has not been explicitly expressed yet. This gap can be filled in by the substitution of  \eqref{eq:dispR} into \specialeqref{eq:k_def}{c}, giving rise to 
\begin{align}\label{eq:dkdt_mid}
    \p_t\bk + (\nabla_k\omega\cdot\nabla)\bk + \nabla\omega = 0,
\end{align}
where $\omega=\omega(\bk(\bx,t),\bx)$ was used. Hence, the material derivative of the wave vector is obtained from \eqref{eq:dkdt_mid} and given by
\begin{align}
   \label{eq:dkdt}
    \dfrac{\rmd \bk}{\rmd t}
    \equiv \p_t\bk + (\nabla_k\omega\cdot\nabla)\bk = - \nabla\omega(\bk,\bx). 
\end{align}

Inserting \specialeqref{eq:raydef}{a,b}, \eqref{eq:bt_exp},  \eqref{eq:d2rdt_exp}, and \eqref{eq:dkdt}  into \eqref{eq:kpdef} and keeping the terms to $\mo(\varepsilon)$ eventually gives rise to the explicit expression of the ray curvature as follows
\begin{align}\label{eq:kp_final}
     \kappa_\approx =
    \left|\dfrac{\p_xU{_2}-\p_yU{_1}}{c_{g,i}}
    + \dfrac{\be_{k,\perp} \cdot\nabla c}{c_{g,i}}
    \right|. 
\end{align}
Here, the subscript `$\approx$' denotes an approximation to the curvature according to \eqref{eq:kpdef}. 
We recall that the local curvature of the rays given by \eqref{eq:kp_final} was derived under the weak current and geometric optics approximations; 
the unit vector $\be_{k,\perp} = [-k_y,k_x]/k$ was defined, which obeys $\be_{k}\cdot\be_{k,\perp}=0$, suggesting that it is always orthogonal to the local wave vectors. We remark that 
the identities $\nabla c=\nabla h\p_hc$ and $c(\bk,\bx)=\sqrt{g\tanh kh(\bx)/k}$ are admitted in \eqref{eq:kp_final} where the terms are all evaluated on the time-dependent rays: $\bx=\br(t)$.  For deepwater waves which admit $\nabla c=\mathbf{0}$, the local curvature of the rays by \eqref{eq:kp_final} recovers to the expressions by \citet{kenyon_wave_1971, dysthe_refraction_2001} and, for the cases in the absence of current, it recovers to 
\citet[their expression (1c)]{arthur_direct_1952}. We also note that the current-gradient term in \eqref{eq:kp_final} can be expressed by the vertical component of the vorticity vector $\zeta = \p_xU{_2} - \p_yU{_1}$.

\section{Limiting cases for wave refraction}\label{sec:evaluation}
In this section, we examine effects of both current and bathymetry on the wave refraction using a few limiting cases. To this end, we firstly assess the accuracy in the analytical approximation given by \eqref{eq:kp_final} by direct numerical simulations presented in \S\ref{sub:model} and next elucidate the underlying novel physics using a few example cases presented in \S\ref{sub:comp}. For the ray path predictions essential for the results presented in this section, we use the open-source ray tracing solver developed by \citet{halsne_ocean_2023}. 
\subsection{Model Validation}
\label{sub:model}
To validate the approximate curvature $\kappa_\approx$ denoted by \eqref{eq:kp_final}, we compare its numerical predictions with those based on the direct numerical implementation of the definition according to \eqref{eq:kpdef}. 
With the ray paths, the curvature based on \eqref{eq:kp_final} can be directly evaluated on the wave rays $\bx = \br(t)$ for the time instants $t = n\Delta t$ with $n\in [0,1,2,...,N_t-1]$, where $\Delta t$ and $N_t$ denote the time interval and total number of time steps used for the numerical results, and $T=\Delta t (N_t-1)$. Similarly, the curvature based on \eqref{eq:kpdef} was readily evaluated applying a second-order accurate central difference scheme on $\br(t)$.

\begin{figure}
    \centering
    \includegraphics[width=\textwidth]{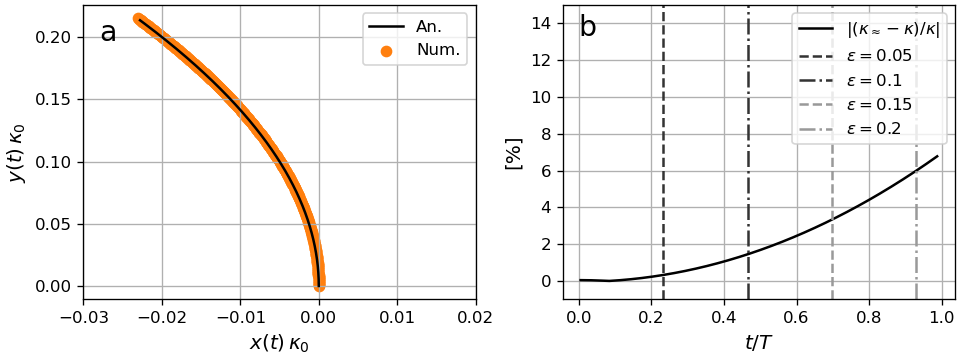}
    \caption{Accuracy in \eqref{eq:kp_final} for a deepwater $6$~s period wave on a shearing current starting at point $\bx_r(0) = $ (0, 0). Panel a) show the analytical ray path (black solid line) and modelled (orange dots) from numerical integration of the ray equations \eqref{eq:raydef} and \eqref{eq:dkdt}. The ray paths are normalized with the initial curvature $\kappa_0$. Indeed, the shape of the normalized ray paths are independent of the shear magnitude and wave period. Panel b) demonstrates the temporal evolution in percentage difference between $\kappa$ and $\kappa_\approx$ for the ray in panel a. Vertical lines denote different values of $\varepsilon = U_1(y)/c_{g,i}$.}
    \label{fig:exact_minus_approx}
\end{figure}

For validating \eqref{eq:kp_final}, we follow \citet{kenyon_wave_1971} as analytical solutions exist for both the ray paths and the material derivative of the wave vector expressed as \eqref{eq:raydef} and \eqref{eq:dkdt}, respectively \citep{longuet-higgins_changes_1961}. In particular, we consider a deepwater wave train on a constant shear current 
\begin{equation}\label{eq:shear_current}
    \bU=[\tau y,0],
\end{equation}
for $y>0$, where $\tau$ is the velocity shear. Inserting the current profile described by \eqref{eq:shear_current} into the approximate curvature leads to $\kappa_\approx = |\tau/c_{g,i}|$, where it has been shown by \citet{kenyon_wave_1971} that this approximation leads to a difference with \eqref{eq:kpdef} by about 10~\% for $\varepsilon \sim$ 0.05. 

The finding reported by \citet{kenyon_wave_1971} is consistent with the comparison shown in Fig.~\ref{fig:exact_minus_approx}. Here, the temporal evolution of wave rays with initial position $\bx_r(0)= $ (0, 0) and initial propagation direction $\theta_0 = \pi/2$ was used, where $\be_k\perp \bu$ is implied. Here and throughout the text, subscript `$0$' refers to initial conditions, i.e., $t=0$. Such rays admit the analytical form $x_r(t) =  -y_r(t)^2 \kappa_0 /2$ \citep{kenyon_wave_1971}, where $\kappa_0 = -2 \tau \Omega_0 / \mathrm{g} $ denotes the initial curvature, and $\Omega_0$ the initial intrinsic frequency. Here, the initial frequency corresponded to a 6~s period wave. Figure~\ref{fig:exact_minus_approx}a demonstrates a good agreement between the ray path predictions by the ray tracing solver and the analytical approximation. As expected, the relative difference between $\kappa_\approx$ and $\kappa$, being defined as $|(\kappa_\approx-\kappa)/\kappa|$, grows in time due to the constant shear $\tau$ and the increase in current speed with $y$ (Fig.~\ref{fig:exact_minus_approx}b); it reaches about 6~\% for $\varepsilon=0.2$, which well falls within the regime as reported by \citet{kenyon_wave_1971}.

\subsection{Wave refraction by both current and bathymetry}
\label{sub:comp}
Our derivations in Section \ref{sec:theory} have demonstrated that the presence of both current and a varying bathymetry can lead to complex interplay, and thereby playing an important role in wave refraction. Roles due to current and a bathymetry can in particular be quantified in a separate manner by noting that the approximate curvature $\kappa_\approx$ expressed as \eqref{eq:kp_final} has the form of linear superposition of current- and bathymetry-altered curvature components denoted by $\kappa_c$ and $\kappa_d$, respectively, where
\begin{equation}\label{eqs:kp_cd}
     \kappa_c = \dfrac{\p_xU_{2}-\p_yU_{1}}{c_{g,i}} 
     \text{ and }
     \kappa_d = \dfrac{\be_{k,\perp} \cdot\nabla c}{c_{g,i}}
     \text{ such that }
     \kappa_\approx = |\kappa_c+\kappa_d|. 
     \specialnumber{a,b,c}
\end{equation}
Here, \specialeqref{eqs:kp_cd}{c} permits us to quantify the individual and combined contribution  of current and depth to the wave refraction, as noted, where both the sign and magnitude of $\kappa_c$ and $\kappa_d$ determine the ultimate effect on the wave refraction. We will use a few limiting cases in the following subsections to explicitly explore the underlying physics. These cases include wave trapping---corresponding to a wave ray which cannot escape---and zero curvature when $\kappa_c=-\kappa_d$. In particular, the wave trapping is typically manifested as total internal reflection, like waves on an opposing current jet or atop an elongated submarine shallow, or as a complete absorption, like waves propagating against a beach. 

\subsubsection{Wave trapping on jet-like currents and complex bathymetry}
\label{subsub:wtrap}
To demonstrate the wave trapping phenomena due to the joint influence of current and depth on the wave refraction, we use the bathymetry profile and jet-like current, being expressed as, respectively, 
\begin{equation}\label{eq:botmount}
    h(x,y) 
    =  \frac{H}{2} \left (1 + \alpha \sin{\left(\pi\frac{x}{L_x} \right)} \cos{\left(\pi\frac{y}{L_y} \right)} \right )
    \text{ and }
    \mathbf{U} = 
    [U_{*} \cos^4{\left( \pi \frac{y}{L_y} - \frac{\pi}{2}\right)} , 0 ]. 
    \specialnumber{a,b}
\end{equation}
Here, $L_x$ and $L_y$ are the characteristic length in the $x$- and $y$-direction, respectively; $H$ denotes the characteristic length in the depth direction; $\alpha \in [0,1]$ denotes the degree of varying bathymetry, leading to the measure of the bathymetry variation in the $x$- and $y$-direction being $\max|\nabla h/h|$. When $\mo (\max|\nabla h/h| / k) \ll 1 $ it corresponds to a slowly varying bathymetry in the horizontal plane, as required in the assumption for the approximate curvature $\kappa_\approx$. Likewise, $U_{*}$ denotes a characteristic current magnitude in the $x$-direction. The corresponding slowly varying current assumption is fulfilled for $\mo (\max|\nabla \mathbf{U}/U| / k) \ll 1 $, where $U = |\mathbf{U}|$ \citep{peregrine_interaction_1976}. Here, and for the subsequent analysis, $L_x=20$ km, $L_y=10$ km, and a wave period of 12~s has been used, unless otherwise stated. The remaining parameters chosen for the numerical predictions are listed in Table~\ref{tab:ambcond}. 

\begin{figure}
    \centering
    \includegraphics[width=\textwidth]{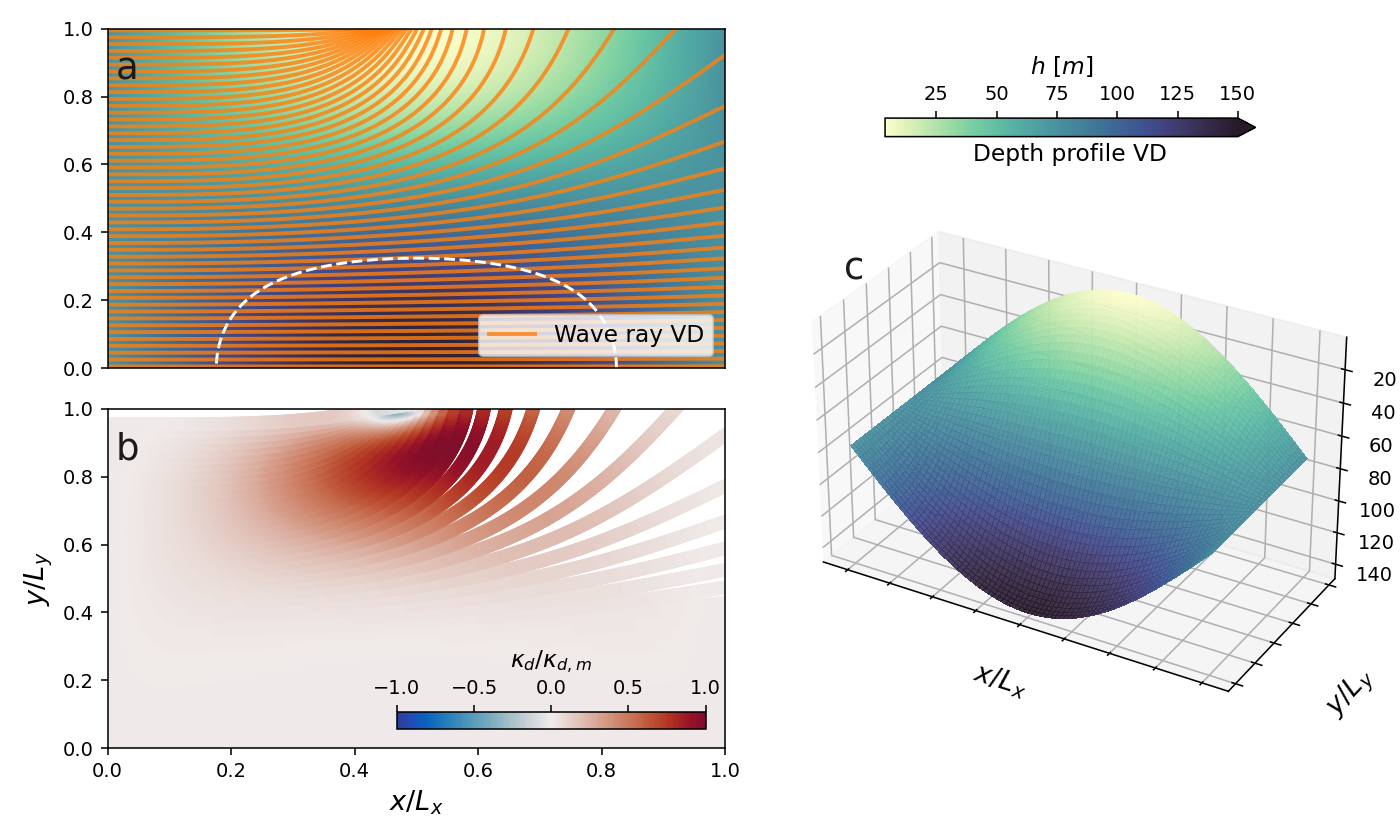}
    \caption{The influence by the variable depth (VD Table~\ref{tab:ambcond}) on the wave propagation. Panel a) show a wave ray with initial period of $12$~s propagating from left to right, with initial propagation direction parallel to the $x$-axis. The bathymetry-altered curvature $\kappa_d$, normalized by its maximum value $\kappa_{d,m}=\max|\kappa_{d}|$, is shown in panel b. The shallowest region in the VD bathymetry simulates a seamount and is shown in panel c, where the ater depth $h$ is given by the background color. White dashed line in panel a denote the $h=\lambda /2$ contour.}
    \label{fig:ray_tracing}
\end{figure}

\begin{table}
  \begin{center}
\def~{\hphantom{0}}
  \begin{tabular}{lccccccc}
      Abbreviation & $f_0$ [Hz]  & $H$ [m]   &   $\alpha$ & $U_{*}$ [m/s] & $k_0 h_0$ 
      & $L_x$ [km] & $L_y$ [km]
      \\[3pt] 
      
       VD  &  0.08  & ~~150 & 0.95 & ~~-~ & ~2 & 20 & 10\\
       DW  &  0.08 & 10 000 & 0 & ~~-~ & $\gg \pi$ & 20 & 10\\
       \hline
       PC  &  0.08 & - & - & ~0.6 & ~2 & 20 & 10\\
       NC  &  0.08 & - & - & --0.6 & ~2 & 20 & 10\\

  \end{tabular}
  \caption{Values for the ambient conditions including varying depths (VD), deep water (DW), positively- (PC) and negatively-oriented current profile (NC) used in the numerical ray tracing simulations in Figures~\ref{fig:ray_tracing}--\ref{fig:raytra_vdC}. Subscript `$0$' refers to initial conditions.}
  \label{tab:ambcond}
  \end{center}
\end{table}

Figure~\ref{fig:ray_tracing} showcases the role of depth-induced refraction on wave propagation, using a variable depth (VD) 
obtained by \specialeqref{eq:botmount}{a} by letting $\alpha=0.95$. For this profile, $h\in [3.75,146.25]
$~m and the shallowest region simulates a seamount. As expected, waves are refracted according to the water-depth gradients, and thus bend towards the shallower regions. The seamount attracts and traps a large portion of the wave rays which are initially located at $y>0.5L_y$. The bathymetry-altered curvature $\kappa_d$ obtains its largest magnitude for the rays being located in the vicinity of the local shallow, but having slightly passed it [see Fig.~\ref{fig:ray_tracing}b around $(0.6L_x,0.8L_y)$].
\begin{figure}
    \centering
    \includegraphics[width=\textwidth]{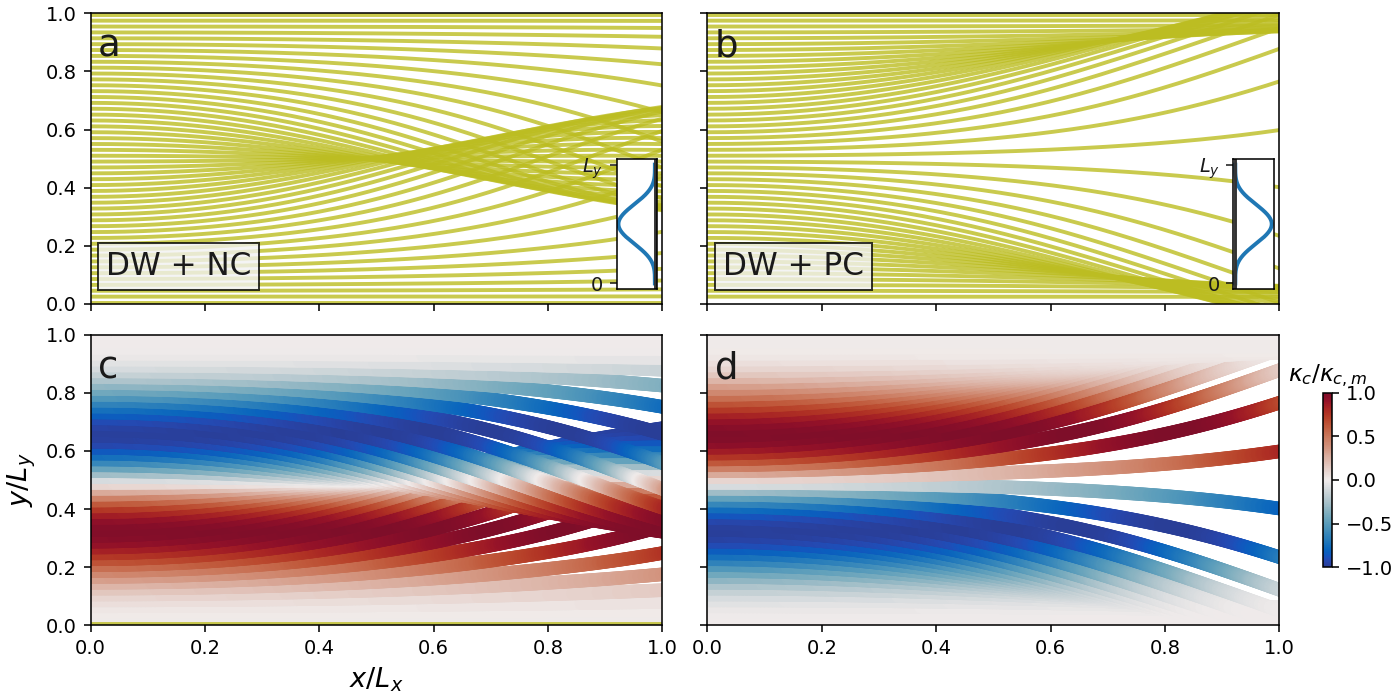}
    \caption{Refraction of deepwater (DW) 12~s period waves atop a negatively- (NC) and positively-oriented current (PC) are shown in panels a and b, respectively. The waves initially propagate from left to right. Inset figures denote the current profiles and their orientation, with further details given in Table~\ref{tab:ambcond}. Panels c and d show the associated current-altered curvature $\kappa_c$ along each wave ray, which is normalized by its maximum value $\kappa_{c,m}=\max|\kappa_{c}|$.}
    \label{fig:raytra_dw}
\end{figure}

Refraction solely due to currents are shown by simulations in deep water (DW) conditions, which include a positively (PC) and negatively oriented current jet (NC). The PC induces caustics at the edges of the jet, while the NC induces caustics at the center of the jet (Figs.~\ref{fig:raytra_dw}a,b). If the domain for the NC case were extended in the $x$-direction, we would have seen the characteristic wave trapping phenomenon at the center of the jet \citep{peregrine_interaction_1976}. In this example, the current-induced curvature $\kappa_c$ is given by
\begin{equation}\label{eq:kc}
    \kappa_c = \dfrac{-2U_{*} \pi }{L_y c_{g,i}} \cos^2\left( \pi \frac{y}{L_y} - \frac{\pi}{2}\right)\sin\left( \pi \frac{2y}{L_y} - \pi\right). 
\end{equation}
Here, the sign of $\kappa_c$ is mirrored about the line $y=0.5L_y$ for the two cases (Figs.~\ref{fig:raytra_dw}c,d). 

\begin{figure}
    \centering
    \includegraphics[width=\textwidth]{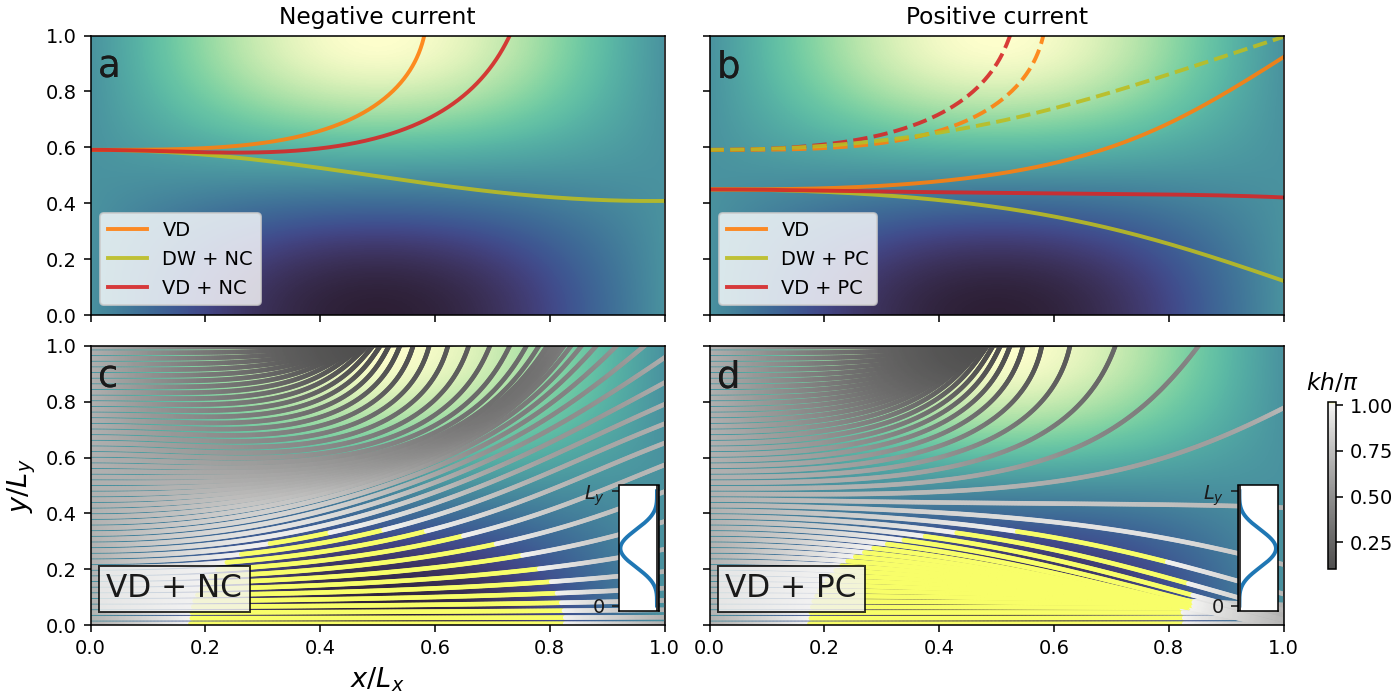}
    \caption{The joint influence by variable depths (VD) and jet-like currents (NC, PC) on wave propagation. Initial wave periods are 12~s. Upper panels show the difference in propagation between depth-only (VD, orange), current-only (yellow) and their joint influence (red), when starting at the same initial position; dashed and solid lines in panel b denote different initial positions. Lower panels show the refraction for several wave rays on the negative current (panel c) and positive current (panel d). The color shading denote $kh/\pi$, where yellow color denote $kh/\pi \ge 1$.} 
    \label{fig:raytra_vdC}
\end{figure}

The joint influence by the bathymetry and current yield more complicated situations (Fig.~\ref{fig:raytra_vdC}). Combining the varying depth and negative current (VD+NC) causes two caustics; one is located at the local shallow and the other takes the form of a meandering tube steered by the bathymetry (Fig.~\ref{fig:raytra_vdC}c). Conversely, when the current is positive, there is a stronger refraction against the local shallow since the following current and bathymetry work together (VD+PC, Fig.~\ref{fig:raytra_vdC}d); this leads to a corresponding decrease of wave rays in the region where $x>0.5L_X$ and $y>0.5L_y$, when compared to the VD-only case (see Fig.~\ref{fig:ray_tracing}a). However, when combining the variable currents and bathymetry, we do recognize the predominant wave ray propagation patterns from the isolated cases; the refraction against the local shallow (Fig.~\ref{fig:ray_tracing}a); the caustic at the center of the opposing jet (Fig.~\ref{fig:raytra_dw}a); and the diverging wave ray pattern at the center of the following jet (Fig.~\ref{fig:raytra_dw}b). 

The combined depth- and current-induced refraction is further highlighted by comparing the ray paths in different cases. Figures~\ref{fig:raytra_vdC}a,b show the ray paths which initiate at the same position  of $(0,0.6L_y)$ but are affected by either the PC, NC, VD, or the combined effects of a current and VD. We see that the effect by the NC is to extend the distance where the ray holds a substantial $k_x$ component compared with the VD solution. As a consequence, the wave ray exits the domain further away from the local shallow, which is illustrated by the red line of Figure~\ref{fig:raytra_vdC}a. The opposite is true for the PC case, where the current gradients refract the wave ray against the VD faster than the bathymetry alone, as is shown by the red dashed line in Figure~\ref{fig:raytra_vdC}b. Thus, the joint effect of the PC and VD is stronger than their individual contribution. 

We also depict in Figure~\ref{fig:raytra_vdC}b the rays with an initial position of $(0,0.44L_y)$. Here, the gradients of the positive current PC almost resemble the gradient of VD, but with different signs, leading to a wave ray with a small curvature, as is illustrated by the solid red line in Figure~\ref{fig:raytra_vdC}b. Such a case will be investigated in more detail in the subsequent subsection. 

We introduce the ratio 
\begin{equation}\label{eq:dominant_scattererer}
  \gamma = {\kappa_d^2  }/{(\kappa_d^2 + \kappa_c^2)},
\end{equation}
such that $\gamma \in [0,1]$, which is referred to as the refraction assessment metric because, when $\kappa_c$ and $\kappa_d$ hold the same sign, it can be used to measure which of the two plays a more dominant role in the wave refraction. Specifically, $\kappa_d$ and $\kappa_c$ dominates for $\gamma>0.5$ and $\gamma<0.5$, respectively. 

\begin{figure}
    \centering
    \includegraphics[width=\textwidth]{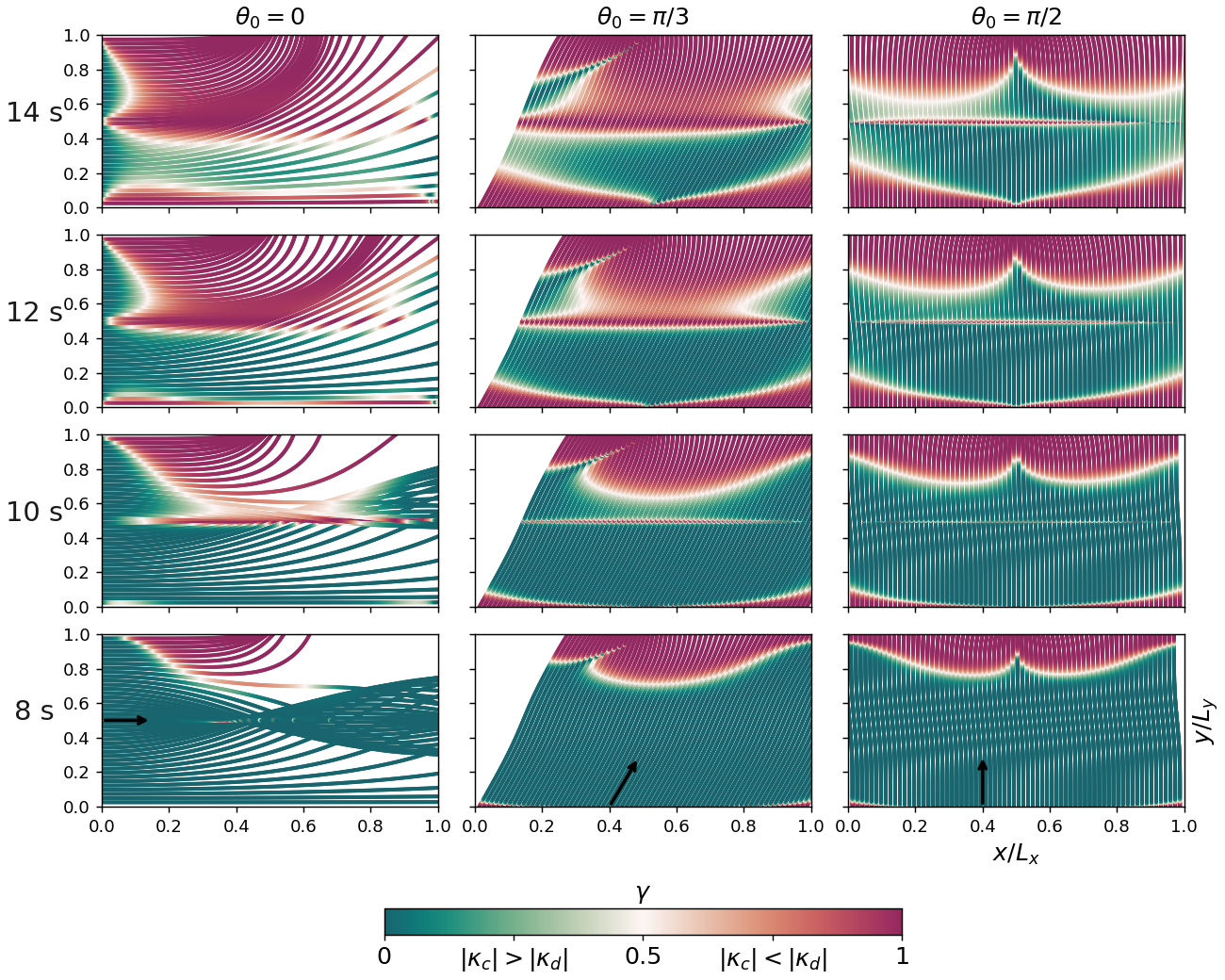}
    \caption{The ray curvature ratio $\gamma$ computed locally along wave rays. Rows from top to bottom show ray tracing simulations for waves with initial periods of 14~s, 12~s, 10~s, and 8~s, respectively. Columns from left to right denote different initial propagation directions $\theta_0$, as indicated by the arrows in the lower row plots. All model simulations have the conditions VD+NC (Table~\ref{tab:ambcond}).}
    \label{fig:dominating_condition}
\end{figure}

The metric \eqref{eq:dominant_scattererer} is shown in Figure~\ref{fig:dominating_condition} for an ensemble of ray tracing simulations, including four different initial wave periods and three initial directions. Here, $\gamma$ is computed locally along the wave rays while propagation atop VD+NC. Besides mapping when and where depth- and current-induced refraction dominate, the results highlight some important physical aspects concerning the two refraction mechanisms. Firstly, Figure~\ref{fig:dominating_condition} demonstrates that the horizontal extent of the bathymetry dominated refraction $\kappa_d$ increases with increasing wave period. This relation is due to the two following reasons: i) $kh$ increases with wave period meaning that longer waves feel more the depth change than shorter ones \citep{kenyon_depth_1983}, and thus scatters accordingly; ii) both terms in $\kappa_\approx$ are inversely proportional to the group velocity $c_{g,i}$, meaning that decreasing wave periods strengthen the contribution from $\kappa_c$. Secondly, Figure~\ref{fig:dominating_condition} highlights the directional dependence of $\kappa_d$; the regions dominated by depth refraction (red-ish colors) have a different horizontal extent when comparing the leftmost and rightmost columns in Figure~\ref{fig:dominating_condition}. The reason for this difference is that the depth gradients $\p_x h \ne \p_y h$, and that the unit vector $\be_{k,\perp} = [-k_y,k_x]/k$ generally holds different values depending on the initial direction. As a consequence, $\kappa_d$ obtains different values. On the contrary, and as seen from \eqref{eq:kc}, $\kappa_c$ does not have such directional dependence.

\subsubsection{Zero curvature} 
\label{subsub:zerokp}
\begin{figure}
    \centering
    \includegraphics[width=\textwidth]{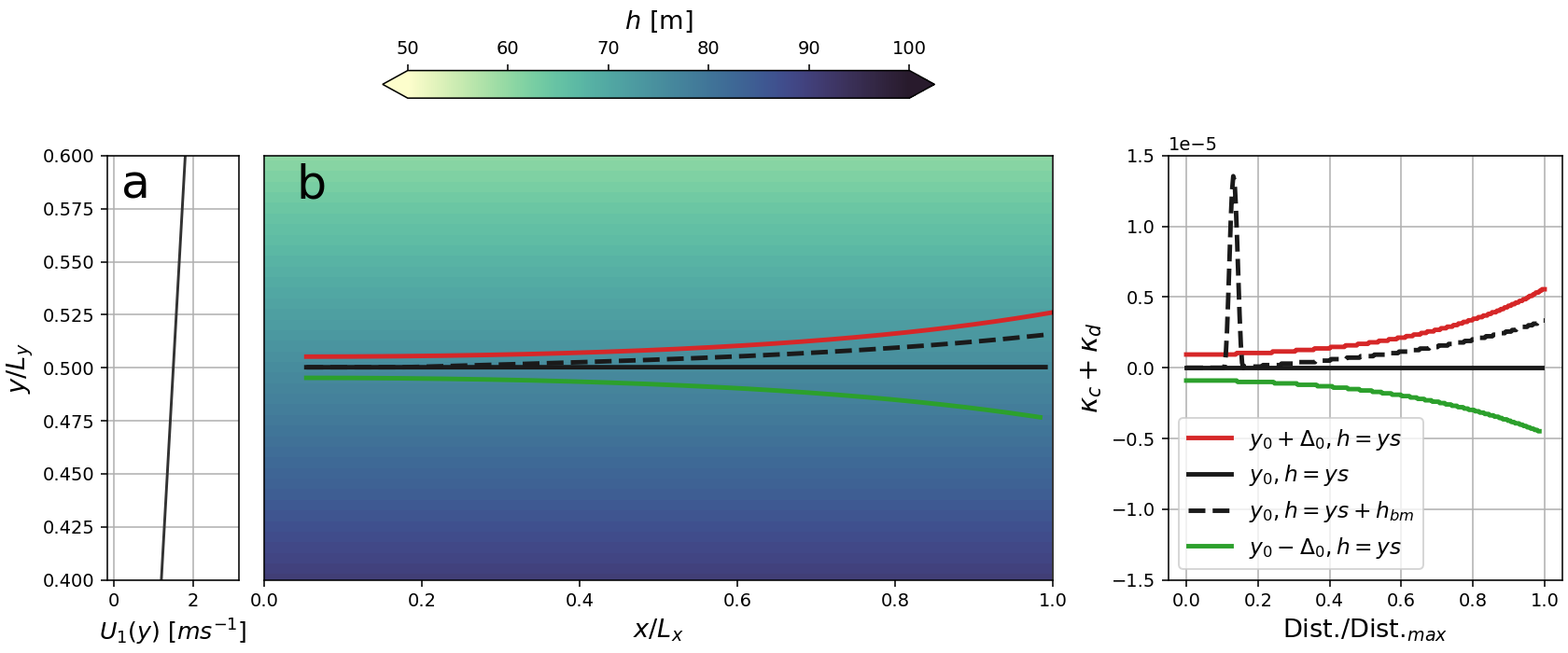}
    \caption{A special case where depth and current refraction equalize and cancel each other. Panel b show wave rays propagating from left to right atop a sloping beach in the $y$-direction and a shear current of type \eqref{eq:shear_current} [see panel a]. The black solid ray, where $\kappa_c = -\kappa_d$, is initially located at $\bx_r(0) = (0.05L_x,y_0)$, where $y_0 = 0.5L_y$. Adjacent rays are perturbed $\pm \Delta_0=\pm 0.005L_y$ in $y$-direction. Panel c show the evolution of $\kappa_c + \kappa_d$ along the wave propagation distance (Dist.). Dashed black line denote a slightly modified twin experiment which adds a small bump in the bathymetry ($h=ys+h_{bm}$) in a subset of the domain, with details outlined in the text.}
    \label{fig:equalizing}
\end{figure}

As highlighted in Figure~\ref{fig:raytra_vdC}b, zero curvature is led when $\kappa_c$ and $\kappa_d$ hold different signs but with a similar magnitude, i.e., $\kappa_c = -\kappa_d$. Such situations thus prevent wave refraction. This is illustrated in Figure~\ref{fig:equalizing} for a prescribed shear current described by \eqref{eq:shear_current} with $\tau = 0.0015$~s\textsuperscript{-1} and a constant sloping beach in the $y$-direction. Here, the depth
\begin{equation}\label{eq:sloping_beach}
    h(x,y) = ys,
\end{equation}
where $s = 0.015$. This example can be used to represent the circumstances where oblique waves enter a gently sloping beach which is subject to longshore currents. To construct a wave ray which recovers $\kappa_c = -\kappa_d$ in the $x$-direction, we solve the implicit equation for wavenumber $k$
\begin{equation}\label{eq:implicit_c}
    \nabla c = \frac{1}{2} \sqrt{\frac{\g k}{\tanh(kh)}} \left[1 - \tanh^2(kh) \right] \nabla h. 
\end{equation}
We consider the initial position $\bx_r(0) = (0.05L_x,y_0)$, where $y_0 = 0.5L_y$ and the water depth $h = 75~m$. Then, \eqref{eq:implicit_c} yields a wave period of 12.7~s, and $k h=1.95$. The resulting wave ray becomes a straight line when the initial propagating direction is parallel to the $x$-axis, and the wave propagates from left to right (Fig.~\ref{fig:equalizing}b). However, small changes in the initial position is decisive for the ray direction. We introduce the small perturbation in position $\Delta_0 = 0.005L_y$, which corresponds to 50~m. The wave ray becomes trapped by the bathymetry if initially located at $y_0 + \Delta_0$, i.e., 50~m closer to shore (red line Fig.~\ref{fig:equalizing}). On the contrary, the ray escapes the bathymetry if initially located $\Delta_0$ further away from shore ($y_0 - \Delta_0$, green line). 

In cases with zero refraction, the wave refraction is very sensitive to any modulation in wavenumber. This is illustrated in a slightly modified twin experiment where the sloping beach is subject to a small perturbation, in the form of a small bump in the bathymetry. Let
\begin{equation}
    h_{\mathrm{bm}} = 10 \sin{ \left(\frac{\pi}{0.05 L_x}(x - 0.15 L_x)\right)} \cos{\left(\frac{\pi}{0.1 L_y} (0.5 + y - 0.45 L_y) \right)},
\end{equation}
represent a local shallow bank, such that $h = ys + h_{\mathrm{bm}}$ in the region $x\in [0.15L_x,0.2L_x]$ and $y\in [0.45L_y,0.55L_y]$. Such a model realization is depiced by the dashed lines in Figure~\ref{fig:equalizing}b and c; while $k$ starts to decrease on the hump, $\nabla c$ starts increasing such that $\kappa_c$ starts dominating. Indeed, the effect of the hump is that the wave ray eventually gets trapped by the bathymetry.

\section{Conclusions}
\label{sec:conclusions}
In this paper, we have examined the complex effects of varying current and depth on the refraction of surface gravity waves. The study is carried out through a newly derived analytical approximation to the wave ray curvature described by \eqref{eq:kp_final} under the assumption of weak current and slowly varying bathymetry. Particularly, the current is assumed to propagate in the same plane as the wave vector, has a depth-uniform velocity profile, and varies slowly in time compared with the phase of the characteristic waves. The approximate curvature recovers to \cite{kenyon_wave_1971, dysthe_refraction_2001} and \cite{arthur_direct_1952} for the cases in deep water and the absence of current, respectively. It is also validated by the open-source ray tracing framework developed by \cite{halsne_ocean_2023}.

The explicit expression of the approximate curvature \eqref{eq:kp_final} is in the form of linear superposition of a current- ($\kappa_c$) and depth-gradient induced component ($\kappa_d$), allowing us to  quantifying the individual and combined contribution  of current and depth to the wave refraction. It indicates that both the sign and magnitude of $\kappa_c$ and $\kappa_d$ play an important role in wave refraction, which have been explicitly explored in a few limiting cases. When $\kappa_c$ and $\kappa_d$ hold the same and opposite sign, the current- and depth-induced components together lead to an enhanced and reduced effect on wave refraction, respectively, compared with their individual contribution. Which of the two plays the dominant role depends on the relative magnitude of $\kappa_c$ and $\kappa_d$. When $\kappa_c\kappa_d>0$, the refraction assessment metric $\gamma$ ($\gamma\in [0, 1]$) described by \eqref{eq:dominant_scattererer} is proposed. For the special cases where $\kappa_c +\kappa_d \approx 0$, we address that additional perturbations due to either current- or depth-induced gradient, even with a small magnitude, can lead to noticeable deviation of the wave propagation direction from its original path in a long distance. 

As noted, the results reported here rely on a theoretical model with explicitly stated assumptions. In other words, the model is limited in its applicability. For instance, it cannot be used to deal with currents whose velocity gradient in the horizontal plane is strong \citep{shrira14}, or currents whose velocity profile is depth dependent \citep{stewart_hf_1974, kirby_surface_1989, ellingsen_approximate_2017, quinn2017explicit,li19}, or water regions of a sudden depth change \citep{trulsen20,li23review}.  Nevertheless, the results from this paper are expected to hold for most practical circumstances in coastal waters and can be readily used in developing new post-processing approaches essential to remote sensing, such as the recovery of current profiles and bathymetry through the measurement of surface waves (see, e.g., \cite{smeltzer19}).

\backsection[Acknowledgements]{T. Halsne would like to thank Dr. Kai H{\aa}kon Christensen and Dr. Mika Petteri Malila for fruitful discussions at the conceptual stage of this work. Y. Li acknowledges the financial support from an European Research Council (ERC) - Starting grant ({\it OceanCoupling}, grant agreement ID: 101164343).Views and opinions expressed are however those of the authors only and do not necessarily reflect those of the European Union or the ERC Executive Agency. Neither the European Union nor the granting authority can be held responsible for them.}


\backsection[Declaration of interests]{The authors report no conflict of interest.}




\bibliographystyle{jfm}
\bibliography{bibliography}

\end{document}